\documentstyle[aps,twocolumn,epsfig]{revtex}
\begin{document}
\newcommand{\ve}[1]{\mbox{\boldmath 1$}}
\twocolumn[\hsize\textwidth\columnwidth\hsize
\csname@twocolumnfalse%
\endcsname

\draft

\title {On the criterion for Bose-Einstein condensation for particles in traps}
\author{C.\ J.\ Pethick$^{1}$ and L.\ P.\ Pitaevskii$^{2,3}$}
\date{\today}
\address{$^1$NORDITA, Blegdamsvej 17, DK-2100 Copenhagen {\O}, Denmark \\
$^2$Department of Physics,
University of Trento,
I-38050 Povo (Trento),
Italy \\
$^3$ Kapitza Institute for Physical Problems, 117334, Moscow, Russia}
\maketitle

\begin{abstract}

We consider the criterion for Bose condensation for particles in a harmonic
trap.  For a fixed angular momentum, the lowest energy state for a cloud of
bosons with attractive interactions is the ground state of the cloud with all
the angular momentum in the center-of-mass motion, and the one-particle
reduced density matrix generally does not have a single large eigenvalue, but
a number of them, suggesting that the state is an example of a fragmented
condensate (Wilkin, Gunn, and Smith, Phys.\ Rev.\ Lett.\ {\bf 80}, 2265
(1998)).  We show that a convenient way to describe correlations in the system
is by defining an internal one-particle reduced density matrix, in which the
center-of-mass motion is eliminated, and that this has a single eigenvalue
equal to the number of particles for the problem considered here.  Our
considerations indicate that care is necessary in formulating a criterion for
Bose-Einstein condensation.

\end{abstract}

\vskip2pc]

In investigations of the properties of the helium liquids, rotation has proved
to be a valuable probe, and this has led to the discovery of vortex lines in
liquid $^4$He and in the superfluid phases of liquid $^3$He.  In the
discussion of superfluidity and Bose-Einstein condensation it is usually
assumed that particles are confined in a box, with a uniform potential
throughout its interior, and with infinitely high walls.  This
situation is very different from that encountered in studies of dilute atomic
vapors in traps, where the potential is generally closer to a
harmonic oscillator one.  We therefore consider the properties of
rotating atomic clouds of bosons at zero temperature.
Among possible ways of putting angular momentum
into the system are the creation of vortex lines \cite{pitaevskii,bp,dalfovo}
and excitation of surface modes and other internal
excitations \cite{stringari}. For traps with an axis
of symmetry, another possibility is to excite the center-of-mass motion, which
for modes having rotational quantum number $m_z=\pm1$ about the symmetry axis
is equivalent to excitation of the simplest surface mode.
The purpose of this paper is to consider the criterion for Bose condensation
in a cloud of particles in a trap.

A criterion for Bose-Einstein condensation was proposed by
Penrose \cite{penrose}, and subsequently elaborated
by Penrose and Onsager
\cite{po}, and by Yang
\cite{yang}.  Consider the one-particle reduced density matrix.  This is
defined in terms of the many-body wave function
$\psi({\bf r}_1, ... {\bf r}_N)$, where the ${\bf r}_i$ are the coordinates
of the particles,  by the equation
\begin{equation}
\rho^{(1)}({\bf r}, {\bf r'}) =
\int d{\bf r}_2 \ldots
d{\bf r}_N
   \psi({\bf r}, {\bf r}_2 \ldots {\bf r}_N)\psi^*({\bf
r}',
{{\bf r}_2}
\ldots {\bf r}_N).
\label{rho}
\end{equation}
The density matrix may be expanded in terms of its eigenfunctions
$\chi_j({\bf r})$ with eigenvalues $\lambda_j$:
\begin{equation}
\rho^{(1)}({\bf r}, {\bf r'}) =   \sum_j\lambda_j  \chi_j^*({\bf
r}')\chi_j({\bf r}).
\end{equation}
The condition generally adopted for the existence of Bose condensation is that
there
should be one eigenvalue that is of order the number of particles, in the
limit when the number of particles tends to infinity.  However, as we shall
indicate below, this can be misleading.

In an instructive paper Wilkin {\em et al.\ }\cite{wilkin} considered
bosons in a harmonic trap, and
showed that for weak attractive forces the lowest state with a given angular
momentum is one in which all the angular momentum is associated with the
center-of-mass motion.  This corresponds to excitation of a dipolar
(magnetic quantum number $m_z=\pm 1$) surface mode of oscillation
\cite{repulsion}. An analysis of the eigenvalues of the one-particle
density matrix revealed that there is not a single large eigenvalue, but
rather there can be many eigenvalues of a comparable size.  This led the
authors to suggest that the state of the rotating system corresponds to that
of a fragmented condensate of the type proposed by  Nozi\`eres and Saint James
\cite{nozieres}.  However, this conclusion seems surprising, because the state
is just the ground state of the cloud, which is Bose-Einstein condensed,
executing center-of-mass motion.  The problem is to identify a quantity that
can reveal the Bose condensation.

Let us consider N spinless bosons in a harmonic trap.
This system has the important property that
the center-of-mass and relative motions of the
particles are
separable, and therefore the wave function may be written in the form
\begin{equation}
\psi({\bf r}_1, ... {\bf r}_N) = \psi_{\rm cm}({\bf R})\psi_{\rm
rel}({\bf q}_1,{\bf q}_2,\ldots, {\bf q}_N),
\label{psi}
\end{equation}
where the center-of-mass coordinate is given by
${\bf R}=(1/N)\Sigma_{i=1}^N{\bf r}_i$ and ${\bf q}_i = {\bf r}_i-{\bf R}$ is 
the coordinate of particle $i$ relative to the center of mass.  Note that the 
definition of the
center-of-mass coordinate implies that only $N-1$ of the ${\bf q}_i$ are 
independent.  Let us now
introduce a quantity which we call the {\it internal}
density matrix
\begin{eqnarray}
\rho_{\rm int}({\bf q}_i, {\bf q}'_i)
&=&\int d{\bf R}
\psi(\{{\bf q}_i+{\bf R}\})\psi^*(\{{\bf q}'_i+{\bf R}\})
\nonumber \\
&=&\psi_{\rm rel}(\{{\bf q}_i\})  \psi^*_{\rm rel}(\{{\bf q}'_i\}).
\label{rhorel}
\end{eqnarray}
This has the property that the center-of-mass motion has been eliminated,
and consequently it depends only on coordinates relative to the center of
mass. One may also construct
reduced density matrices in terms of the relative
coordinates in the standard way.  For example, the internal one-particle
reduced density matrix is
\begin{equation}
\rho^{(1)}_{\rm int}({\bf q}_1, {\bf q}'_1) =
 \int d{\bf q}_2
... d{\bf q}_{N-1}   \rho_{\rm int}({\bf q}_i, {\bf q'}_i).
\label{rhorelone}
\end{equation}
Consequently the internal density matrix for the lowest energy state of a
cloud of bosons with weak attractive interactions and nonzero angular
momentum in a harmonic trap is the same as
for the ground state in the absence of rotation.
It has one large eigenvalue
which is equal to $N$, to within terms of unity, the deviation from $N$
being due to the constraint that the sum of the coordinates relative to the
center of mass must vanish.

Our discussion indicates that caution is required in formulating a criterion
for Bose-Einstein condensation in finite systems, and that the one generally
used can be misleading.  For systems in harmonic traps the internal density
matrix introduced above can be a useful concept, because it removes the
effects of the center-of-mass motion which, for a rotating cloud, tends to
spread out the weight of the usual one-particle reduced density matrix over a
number of states.

We are grateful to Ben Mottelson for helpful conversations.

\end{document}